\documentclass[epj]{svjour}
\usepackage{graphics}
\begin{document}
\newcommand{\oss}{$1~^1\rm{S}$}
\newcommand{\tts}{$2~^3\rm{S}$}
\newcommand{\tss}{$2~^1\rm{S}$}
\newcommand{\ttp}{$2~^3\rm{P}$}
\newcommand{\tsp}{$2~^1\rm{P}$}
\newcommand{\thts}{$3~^3\rm{S}$}
\newcommand{\thss}{$3~^1\rm{S}$}
\newcommand{\thtp}{$3~^3\rm{P}$}
\newcommand{\thsp}{$3~^1\rm{P}$}
\newcommand{\transIR}{$2~^3\rm{S}_1 \rightarrow 2~^3\rm{P}_2$}
\newcommand{\transUV}{$2~^3\rm{S}_1 \rightarrow 3~^3\rm{P}_2$}
\newcommand{\molgsgs}{$2~^3\rm{S}+2~^3\rm{S}$}
\newcommand{\molgsgp}{$2~^3\rm{S}+2~^3\rm{P}$}
\newcommand{\molgsep}{$2~^3\rm{S}+3~^3\rm{P}$}
\newcommand{\molgses}{$2~^3\rm{S}+3~^3\rm{S}$}
\newcommand{\molsingletgsgs}{$2~^1\rm{S}+2~^3\rm{S}$}
\newcommand{\Dtwo}{$2~^3\rm{S}_1+2~^3\rm{P}_2$}
\newcommand{\Done}{$2~^3\rm{S}_1+2~^3\rm{P}_0$}
\newcommand{\qsgp}{$^5\Sigma_g^+$}
\newcommand{\bra}[1]{%
    \langle #1|}
\newcommand{\ket}[1]{%
    | #1 \rangle}
\newcommand{\bracket}[2]{%
    \langle #1 | #2 \rangle}
\newcommand{\matrixel}[3]{%
    \langle #1 | #2 | #3 \rangle}
\newcommand{\ts}[1]{%
    _{\rm{#1}}}

\title{Prospects for measurement and control of the scattering length of metastable helium using
photoassociation techniques}
\author{J. C. J. Koelemeij\inst{1}\thanks{\emph{Present address:} Time and Frequency Division,
National Institute of Standards and Technology, 325 Broadway, Boulder,
Colorado 80305}
and M. Leduc\inst{1}
}                     
\authorrunning{Koelemeij and Leduc}
\titlerunning{Prospects for measurement and control of the scattering length of metastable helium..}

%
\institute{\'Ecole Normale Sup\'erieure and Coll\`ege de France,
Laboratoire Kastler Brossel, 24 rue Lhomond, 75231 Paris CEDEX 05,
France}
\date{Received: date / Revised version: date}
%
\abstract{A numerical investigation of two-laser photoassociation
(PA) spectroscopy on spin-polarized metastable helium (He*) atoms
is presented within the context of experimental observation of the
least-bound energy level in the scattering potential and
subsequent determination of the $s$-wave scattering length.
Starting out from the model developed by Bohn and Julienne [Phys.
Rev. A \textbf{60}, (1999) 414], PA rate coefficients are obtained
as a function of the parameters of the two lasers. The rate
coefficients are used to simulate one- and two-laser PA spectra.
The results demonstrate the feasibility of a spectroscopic
determination of the binding energy of the least-bound level. The
simulated spectra may be used as a guideline when designing such
an experiment, whereas the model may also be employed for fitting
experimentally obtained PA spectra. In addition, the prospects for
substantial modification of the He* scattering length by means of
optical Feshbach resonances are considered. Several experimental
issues relating to the numerical investigation presented here are
discussed.
} 
\maketitle
\section{\label{introduction}Introduction}
Within the context of ultracold quantum gases, Bose-Ein\-stein
condensation (BEC) of metastable helium has opened up the way for
experiments exploiting the 19.8~eV internal energy of the He*
atoms in novel detection
schemes~\cite{Robert2001,Sirjean2002,Seidelin2003}. The accuracy
of the results obtained in contemporary experiments, however, is
compromised by the lack of an accurate determination of the value
of the He* $s$-wave scattering
length~\cite{Gadea2002,Santos2001,Leduc2002,Seidelin2004,Tol2004}.
Another limitation in ultracold He* experiments is the absence of
magnetic Feshbach resonances to control the interaction between
the atoms, which is a consequence of the zero nuclear spin of
$^4$He.

\begin{figure}
\resizebox{0.75\columnwidth}{!}{%
  \includegraphics{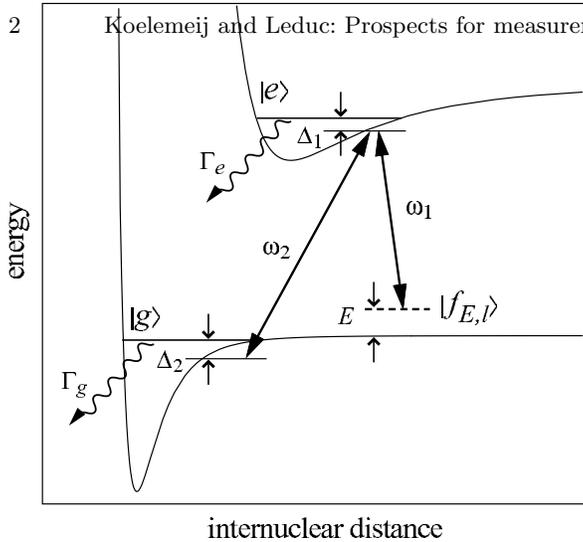}
}
\caption{\label{fig:PAschematic}Schematic view of two-laser
photoassociation, employed here to convert a pair of free atoms
(denoted as $\ket{f_{E,l}}$) into the least bound molecular state
$\ket{g}$ supported by the scattering potential.}
\end{figure}

Standard techniques based on photoassociation (PA) spectroscopy
have enabled accurate determination of the scattering length for
most alkali elements. For the experimental determination of the
scattering length two widely-used methods exist. The first one
relies on determining the binding energy of the least-bound level
in the scattering potential via a stimulated two-photon Raman
transition, from which the scattering length can be inferred (see,
for instance, Refs.~\cite{Abraham1995,Wang2000}). This method is
schematically depicted in Fig.~\ref{fig:PAschematic}. In the
second method, Franck-Condon oscillations observed in the
linestrength of transitions from the free-particle ground state to
vibrational levels in an electronically excited molecular state
are used to determine the nodes in the scattering wavefunction,
from which the scattering length may also be
deduced~\cite{Heinzen}.

For the case of two spin-polarized $2~^3S_1$ helium atoms, elastic
scattering takes place in a \qsgp\ potential. However, the
application of standard techniques to locate the least-bound state
in this potential is not entirely straightforward. The situation
is complicated by the autoionization of metastable helium dimers,
which occurs at short internuclear range. Only dimers in a pure
quintet spin configuration at short range will experience
suppression of autoionization such that well-resolved vibrational
resonances exist. This reduces \emph{a priori} the number of
useful pathways for a PA experiment. In addition, the short-lived
singlet and triplet vibrational states effectively merge into a
continuum which may result in unwanted background signal during PA
experiments. Indeed, long-lived vibrational levels have recently
been observed in molecular states which possess quintet character
at short internuclear range~\cite{Leonard2004,Kim2004}. Some of
these states, which are located below the $D_2$ (\Dtwo) asymptote,
might be useful intermediate levels for a two-laser PA experiment
aiming at the determination of the energy of the least-bound level
in the scattering potential.

An interesting alternative is offered by vibrational levels in the
purely long-range $0_u^+$ potential well below $D_0$ (\Done),
observed by L\'eonard \textit{et
al.}~\cite{Leonard2003a,Leonard2003b,Kim2004}; see also
Fig.~\ref{fig:PA}. For such states the classical inner turning
point lies at long range, which leads to total inhibition of the
short-range autoionization process. Despite their long-range
location, the vibrational wavefunctions can be estimated to have
sufficient Franck-Condon (FC) overlap with the least-bound
wavefunction in the \qsgp\ potential. The \qsgp\ potential has
been calculated \textit{ab initio} by St\"arck and
Meyer~\cite{Starck1994}, from which a scattering length of
150~a$_0$ is inferred. Gadea \textit{et al.} reconsidered the same
potential and concluded that the least-bound vibrational level
must be $v=14$, with a binding energy of at least
$-10$~MHz~\cite{Gadea2002}. This translates to a scattering length
smaller than 340~a$_0$. By comparison, the most accurate
experimental value for the He* scattering length to date (which
was obtained by observing inelastic collisions at BEC threshold)
is $195^{+40}_{-20}$~a$_0$~\cite{Seidelin2004}, in agreement with
the theoretical prediction.

Control of the value of the scattering length may be achieved by
means of an \emph{optical} Feshbach resonance (OFR), as first
proposed by Fedichev \textit{et al.}~\cite{Fedichev1996a}, and
recently observed by Theis \textit{et al.}~\cite{Theis2004}. It
was pointed out that a laser coupling a colliding pair of
ultracold atoms to a bound level in an excited molecular state may
cause a significant modification of the scattering length. This
scheme, however, also implies loss of trapped atoms after
spontaneous decay of the excited state. A suitable combination of
experimental parameters should be chosen to minimize these losses.

Both two-laser PA and optical Feshbach resonances can be
investigated numerically using a scattering model derived within
the framework of generalized multichannel quantum defect theory,
as done by Bohn and Julienne~\cite{Bohn1999}. Their results, which
are recapitulated in Section~\ref{theorybgr}, include scattering
matrix elements and a recipe for calculating two-laser PA rate
coefficients. In Section~\ref{res2PA}, two-laser PA spectra are
simulated for the circumstances pertaining to the experiment
described in Refs.~\cite{Leonard2003a,Leonard2003b}. Next, in
Section~\ref{numresOFR}, an account of the achievable modification
of the He* scattering length using an OFR is given. In addition,
experimental issues which may play a role in the choice of a
suitable vibrational level for the OFR are discussed. Conclusions
are presented in Section~\ref{conclusion}.
\section{\label{theorybgr}Theoretical background}
\subsection{\label{PAtheory}Two-laser photoassociation}
Experimentally, PA spectra are obtained by monitoring the
production rate of a certain product $P$ resulting from the decay
of the level being populated, while scanning the frequency of the
PA laser. For instance, $P$ could be the photon emitted during
spontaneous decay. Plotting the rate coefficient associated with
the production of $P$, $K_P$, versus PA laser frequency therefore
simulates the PA spectrum. For a two-laser PA experiment involving
a collection of atoms characterized by a Maxwell-Boltzmann energy
distribution, $K_P$ can be expressed as ~\cite{Napolitano1994}
\begin{eqnarray}\label{eq:Kp}
\lefteqn{K_P(T,\omega_1,\omega_2,I_1,I_2)=(2\pi\hbar Q_T)^{-1}\sum_{l=0}^{\infty}(2l+1)} \nonumber \\
&&\times\int_{0}^{\infty}{\left|S_P(E,l,\omega_1,\omega_2,I_1,I_2)\right|^2e^{-E/k_BT}dE}.
\nonumber \\
\end{eqnarray}
Here, the integration is over all initial scattering energies $E$.
$\left|S_P(E,l,\omega_1,\omega_2,I_1,I_2)\right|^2$ is the squared
absolute value of the $S$-matrix element, which represents the
probability of production of $P$ as a function of the frequencies
$\omega_1$, $\omega_2$ and intensities $I_1,I_2$ of lasers 1 and
2, respectively. The summation index $l$ labels the partial wave
contributing to the scattering process, and $Q_T=(\mu
k_BT/2\pi\hbar^2)^{3/2}$ is the translational partition function
with $\mu$ denoting the reduced mass of the system. For the
two-laser case, there are three possibilities for the final state
after the scattering process. These correspond to the transition
of the initial free-atom pair (denoted as an energy-normalized
state $\ket{f_{E,l}}$) to the excited state, $\ket{e}$, the bound
molecular ground state, $\ket{g}$, or back to $\ket{f}$ (see also
Fig.~\ref{fig:PAschematic}). For the case of PA of He*, the term
"ground state" refers to the metastable (\molgsgs) \qsgp\ state,
and not the actual ground state of the helium dimer. The
probability for each of these transitions to take place is
determined by the squared absolute value of the corresponding
$S$-matrix element, the explicit form of which is given in
Eqs.~(4.8--4.11) of Ref.~\cite{Bohn1999}. The $S$-matrix elements
themselves depend on the rate $\gamma\ts{s}$ of stimulated
free-to-bound transitions induced by laser 1, the bound-bound Rabi
coupling by laser 2, $\Omega$, the decay widths $\Gamma_e$ and
$\Gamma_g$ of the excited and ground molecular bound states,
respectively, and the detunings $\Delta_1$ and $\Delta_2$ of
lasers 1 and 2. The detunings are defined as
$\Delta_1=E_e/\hbar-\omega_1$ and
$\Delta_2=E_g/\hbar-(\omega_1-\omega_2)$, where $E_e$ and $E_g$
are the energies (with respect to the \molgsgs\ asymptote) of the
excited-state and ground-state vibrational levels, respectively.
This definition of the detuning implies that a negative value for
$\Delta_i$ corresponds to a \emph{blue} detuning. The stimulated
rate $\gamma\ts{s}$ is obtained from Fermi's golden rule:
\begin{equation}\label{eq:gamma}
\gamma\ts{s}=(2\pi/\hbar)\left|\matrixel{f_{E,l}}{V\ts{rad}}{e}\right|^2,
\end{equation}
where the radiative coupling, $V\ts{rad}$, by laser 1 is given by
\begin{equation}\label{eq:Vrad}
V\ts{rad}=d\left(\frac{I_1}{\epsilon_0 c}\right)^{1/2}.
\end{equation}
Here, $d$ is the \emph{atomic} dipole moment operator, and the
polarization of the light is ignored for the moment. It should be
mentioned that the use of energy-normalized states automatically
includes the correct Wigner-law threshold behaviour, which means
that $\gamma\ts{s}$ vanishes as $E^{(2l+1)/2}$ for
$E\rightarrow0$.

$V\ts{rad}$ also determines the Rabi coupling between the ground
and excited bound states, and the Rabi frequency, $\Omega$, is
given by
\begin{equation}\label{eq:Rabi}
\Omega=\matrixel{e}{V\ts{rad}}{g}/\hbar.
\end{equation}
\subsection{\label{opticalFeshbachtheory}Optical Feshbach resonance}
To create an optical Feshbach resonance, only one PA laser is
required. The $S$-matrix elements for two-laser PA may also be
used to describe PA by a single laser by setting the intensity of
laser 2 to zero. Then, from the scattering matrix element $S_{00}$
for elastic scattering (see Eqs.~(4.9--4.11) in
Ref.~\cite{Bohn1999}) a complex-valued phase shift, $\delta$, may
be obtained using the relation~\cite{Bohn1997}
\begin{equation}\label{eq:complexphaseshift}
S_{00}=\exp\left(2i\delta\right).
\end{equation}
For low scattering energies, this complex phase shift translates
to a complex-valued scattering length, $a$, according to
\begin{equation}\label{eq:complexscatteringlength}
a=A-iB\equiv-\lim_{k\rightarrow0}\frac{1}{k}\tan\delta.
\end{equation}
Here, $k=\sqrt{2\mu E/\hbar^2}$ denotes the initial wave number.
Using the explicit expression for $S_{00}$ given in
Ref.~\cite{Bohn1999}, it can be shown that
\begin{equation}\label{eq:A}
A=a\ts{nat}+\frac{1}{2k}\frac{\gamma\ts{s}[E-\hbar\Delta_1-E(I_1)]}{[E-\hbar\Delta_1-E(I_1)]^2+(\Gamma_e/2)^2},
\end{equation}
where $a\ts{nat}$ is the natural value of the scattering length,
and
\begin{equation}\label{eq:B}
B=\frac{1}{4k}\frac{\gamma\ts{s}\Gamma_e}{[E-\hbar\Delta_1-E(I_1)]^2+(\Gamma_e/2)^2}.
\end{equation}
Here, $E(I_1)$ is an energy- and intensity-dependent frequency
shift~\cite{Bohn1999}, which will be ignored below as it does not
fundamentally change the results. The imaginary part of $a$
accounts for losses due to spontaneous emission occurring in the
excited state. By tuning the intensity and/or the detuning from
the resonance, $a$ may be varied. It should be noted that within
this picture, mixing due to state dressing of the initial free
state and the excited bound state is ignored. This has been
predicted to give rise to additional frequency shifts and even
extra resonances for PA at relatively high
intensity~\cite{Montalvao2001,Simoni2002}.

The real and imaginary parts of $a$ can be converted into rate
coefficients, $K\ts{el}$ and $K\ts{inel}$, for elastic and
inelastic collisions, respectively~\cite{Bohn1997}:
\begin{equation}\label{eq:Kel}
K\ts{el}=\frac{8\pi\hbar}{\mu}k(A^2+B^2),~~K\ts{inel}=\frac{8\pi\hbar}{\mu}B.
\end{equation}
Here, $K\ts{inel}$ relates to scattering events in which two atoms
are lost (in Section~\ref{numresOFR} the various loss processes
will be described).
%
\section{\label{numres}Numerical results}
\subsection{\label{res2PA}Two-laser photoassociation spectra}
\begin{figure}
\resizebox{1\columnwidth}{!}{%
  \includegraphics{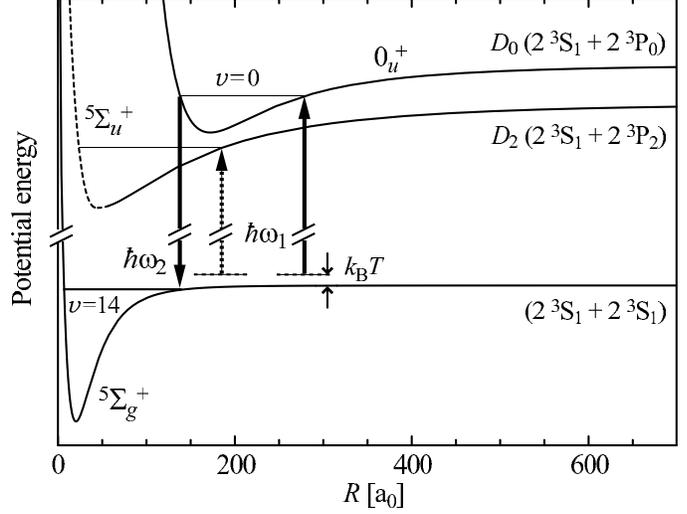}
}
\caption{\label{fig:PA}PA of spin-polarized He* atoms. Laser 1
(dashed arrow) can couple two free atoms (which are initially in
the $^5\Sigma_g^+$ scattering potential) to an excited bound
state, for instance in the potential correlating to the $D_2$
asymptote, thereby creating an OFR. The dashed short-range part of
the potential well indicates that it is not accurately known.
Laser 1 may also be tuned to $v=0$ vibrational state in the purely
long range $0_u^+$ potential below the $D_0$ asymptote (solid
arrow). Laser 2 may then resonantly couple this state to the
least-bound ($v=14$) vibrational state in the $^5\Sigma_g^+$
scattering potential. The difference frequency of the two lasers
is directly related to the binding energy of the $v=14$ state.}
\end{figure}%
As has been pointed out in Ref.~\cite{Leonard2003a}, the $v=0,J=1$
state in the $0_u^+$ potential well below $D_0$
(Fig.~\ref{fig:PA}) is expected to have sufficient FC overlap with
the least-bound state in the $^5\Sigma_g^+$ scattering potential.
The value of the squared FC-overlap integral is of the order of
$10^{-1}$ (which is more than one order of magnitude larger than
for any of the other five bound states observed in the $0_u^+$
well), for which case the efficiency of the second step in the
two-laser PA process should be reasonable. The fact that the
$v=0,J=1$ line could be observed with a good signal-to-noise (S/N)
ratio indicates that also the the first step can be made at a
sufficiently high rate~\cite{Leonard2003a}. For the simulation of
two-laser PA, the $v=0,J=1$ state will be used as the intermediate
state, and conditions similar to those in the experiments
described in Refs.~\cite{Leonard2003a,Leonard2003b} will be
assumed. These imply PA on a magnetostatically trapped cloud
consisting of $10^5$--$10^6$ spin-polarized He* atoms, at
densities $10^{12}$--$10^{13}$~cm$^{-3}$, and temperatures
$10$--$100~\mu$K. It was checked that for such low temperatures,
all $l\ge 2$ contributions to the summation in Eq.~(\ref{eq:Kp})
are negligibly small as compared to to the $s$-wave contribution.
Since odd partial waves do not contribute to the scattering of
identical bosonic He* atoms, only the $l=0$ term is retained. This
is consistent with the fact that no vibrational levels in $0_u^+$
with total rotational quantum number $J>1$ were observed
experimentally~\cite{Leonard2003a,Leonard2003b}.

For the evaluation of the matrix elements in Eqs.~(\ref{eq:gamma})
and (\ref{eq:Rabi}), the wavefunctions of the involved states must
be known. For the $^5\Sigma_g^+$ wavefunctions, these are obtained
by numerically solving the radial Schr\"odinger equation using the
potential data from Ref.~\cite{Starck1994}. Of course, the
wavefunctions depend on the actual shape of the potential, which
is not accurately known. For the simulation of two-laser PA
spectra the above mentioned potential data are used. In case that
the actual potential turns out to deviate from the one used here,
the difference between the two potentials will predominantly
affect the radiative coupling (Eq.~(\ref{eq:Vrad})) via the FC
overlap of the radial wavefunctions (see also the Appendix).
However, for most potential curves lying within the range of the
theoretical prediction, the differences due to this effect will
not be too large. Also the $0_u^+$ wavefunction is found by
numerical integration of the Schr\"odinger equation, using the
Born-Oppenheimer potential obtained from diagonalization of the
long-range interaction between a $2~^3$S and a $2~^3$P atom. Here,
the long-range interaction includes the electrostatic resonant
dipole-dipole interaction (which scales with the internuclear
distance $R$ as $R^{-3}$), the Van der Waals interaction (scaling
as $R^{-6}$), and the atomic relativistic interactions
(spin-orbit, spin-spin etc.)~\cite{Leonard2003b}. Further details
of the evaluation of the matrix elements are postponed to
Appendix~B.
\begin{figure}
\resizebox{0.87\columnwidth}{!}{%
  \includegraphics{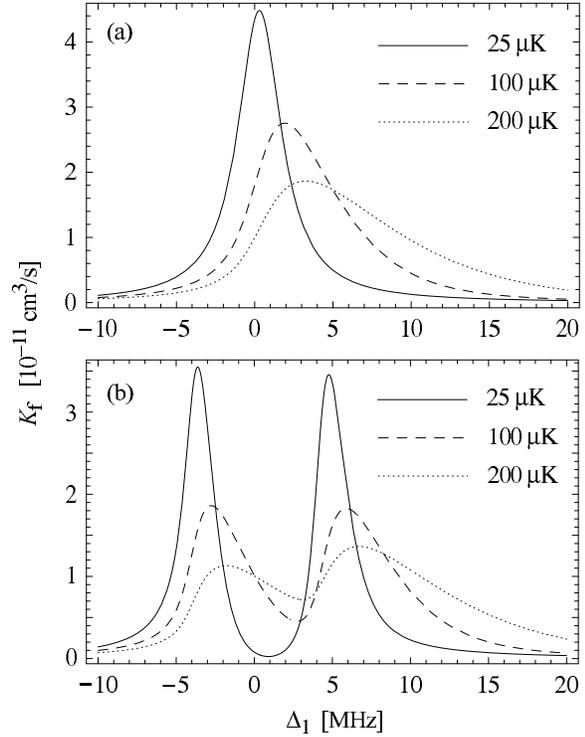}
}
\caption{\label{fig:PA1}Rate coefficients for observation of
spontaneous emission from the excited state plotted as a function
of $\Delta_1$, for various temperatures. (a) Single-laser
lineshapes with $I_1=100~I\ts{sat}$. (b) Two-laser lineshapes,
with $\Delta_2=0$, $I_1=100~I\ts{sat}$ and $I_2=400~I\ts{sat}$.
The spectra in (b) correspond to the case of \emph{frustrated
PA}.}
\end{figure}

Having acquired all necessary information, Eq.~(\ref{eq:Kp}) can
be employed to calculate the rate coefficient $K\ts{f}$ for
observation of fluorescence decay of the excited state. In
Fig.~\ref{fig:PA1}(a), $K\ts{f}$ is plotted versus laser-frequency
detuning for a single-laser PA experiment, and for various
temperatures. For the 25~$\mu$K case, the line profile has a width
close to the spontaneous decay width of $\sim3$~MHz, which is
about twice the width of the atomic \transIR\
line~\cite{Leonard2003b}. For higher temperatures the line becomes
increasingly more shifted, broadened and skewed. It also follows
that for intensities well below $10^4~I\ts{sat}$ (with
$I\ts{sat}=1.67$~W/m$^2$ the saturation intensity of the atomic
\transIR\ transition), broadening due to saturation does not play
a significant role. The bandwidth of the PA laser is neglected,
which is justified for a laser with a few-hundred kHz bandwidth.

\begin{figure}
\resizebox{1\columnwidth}{!}{%
  \includegraphics{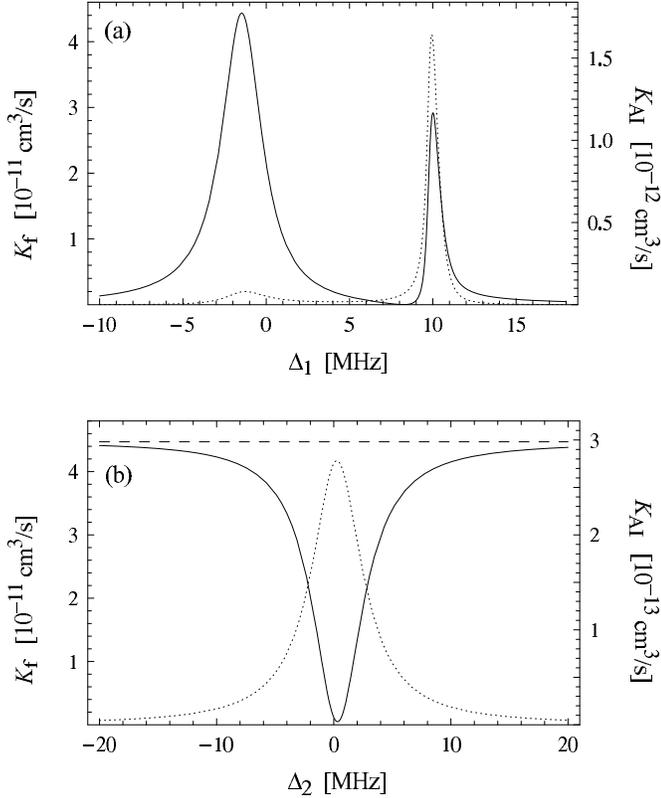}
}
\caption{\label{fig:PA2}Rate coefficients for observation of
fluorescence decay($K\ts{f}$, solid curve) and for observation of
autoionization of \molgsgs\ molecules formed by stimulated
two-photon transitions ($K\ts{AI}$, dotted curve). (a) Two-laser
lineshapes with $I_1=100~I\ts{sat}$, $I_2=400~I\ts{sat}$, and
$\Delta_2=8$~MHz. (b) Frustrated PA with $I_1=I_2=100~I\ts{sat}$.
Here, $\Delta_2$ is scanned while $\Delta_1\approx 0$
(corresponding to the peak of the single-laser line, whose height
is indicated by the dashed horizontal line). The temperature is
25~$\mu$K for all plots}
\end{figure}

For the two-laser case depicted in Fig.~\ref{fig:PA1}(b), laser 1
is scanned across the Autler-Townes doublet induced by laser 2.
Again, at higher temperatures the spectrum becomes more shifted,
and the Autler-Townes doublet becomes less resolved. In
Fig.~\ref{fig:PA2}(a), an example of stimulated two-photon Raman
PA is depicted. Laser 2 is now detuned several linewidths below
resonance, and scanning laser 1 from blue to red shows the
"unperturbed" line and the narrow Raman line, respectively. The
width of the latter is determined by the autoionization width of
the \qsgp\ state (see below), and by thermal broadening. An
alternative approach for detecting the two-photon transition is
shown in Fig.~\ref{fig:PA2}(b). Here, laser 1 is kept at the
maximum of the single-laser line (i.e. $\Delta_1\approx 0$),
whereas laser 2 is scanned. As long as laser 2 is not resonant, a
background signal due to PA will be present. When laser 2 becomes
resonant, the single-laser line will split up into the
Autler-Townes doublet, causing a reduction of PA signal. This
"frustrated PA" scheme may be convenient for an initial search, as
the decrease in PA signal can be made 100\%.

At this point, the connection with the experiment described in
Refs.~\cite{Leonard2003a,Leonard2003b} demands some attention. In
this experiment, the PA spectra were obtained in a magnetic trap
using a novel calorimetric detection scheme, which is briefly
outlined below. Spontaneous decay of the excited state will
produce fast atom pairs with energies up to $\sim 30$~mK, with
some of the pairs being in trapped magnetic substates. If the
temperature of such atom pairs is below the 10~mK trap depth,
their kinetic energy will be deposited in the trapped cloud by
rethermalizing collisions. For a sufficiently cold cloud, this
leads to substantial heating due to resonant PA, while the
decrease in atom number due to the escaping atom pairs may also be
detected. Since the heating is related to spontaneous decay of the
excited state, the spectra plotted in Figs.~\ref{fig:PA1} may also
resemble those measured using the calorimetric method. However,
although it can be estimated that the above mechanism nearly
explains all the heating in the one-laser PA
experiments~\cite{thesisLeonard}, other mechanisms should be
considered too. Spontaneous decay of the excited state may also
result in a bound $^5\Sigma_g^+$ state, which will eventually
decay by autoionization. Due to its small binding energy, the
radial wavefunction of the $v=14$ vibrational state extends far
enough to have reasonable FC overlap with the purely long range
vibrational states in $0_u^+$, which have classical inner turning
points near 150~a$_0$. By contrast, the outer turning points of
the vibrational states in \qsgp\ with $v\le13$ lie at internuclear
distances smaller than $30$~a$_0$~\cite{Starck1994}. Therefore,
the $v=14$ state is most likely to be populated via spontaneous
decay. Combining a semiclassical estimate of the $v=14$ vibration
frequency with the Penning ionization suppression factor given in
Ref.~\cite{Fedichev1996b}, one arrives at an approximate value for
the inverse autoionization-limited lifetime
$\Gamma_g=(5~\mu\rm{s})^{-1}$. The autoionization will produce
fast He$^+$ or He$_2^+$ ions, which may cause heating via
collisions with trapped He* atoms similar to the heating observed
in a He* BEC~\cite{Santos2001}. This process may become important
in the case of stimulated production of $^5\Sigma_u^+$ molecules.
To illustrate this, Figs.~\ref{fig:PA2} display the rate
coefficients $K\ts{AI}$ for decay by autoionization. Heating due
to $K\ts{AI}$ should be accounted for by multiplying $K\ts{AI}$
with a certain constant and adding it to $K\ts{f}$ to simulate the
calorimetric spectrum. Likewise, for the frustrated PA spectrum
depicted in Fig.~\ref{fig:PA2}(b), the heating due to
$^5\Sigma_g^+$ bound state formation will shallow the dip near
$\Delta_2=0$. However, as $K\ts{AI}$ for this case is more than
two orders of magnitude smaller than $K\ts{f}$, this effect might
be relatively small. The lineshapes for $K\ts{AI}$ in
Figs.~\ref{fig:PA2} also suggest the interesting option of ion
detection for observing the two-laser transition. Finally, heating
due to off-resonant atomic scattering should be taken into
account. For intensities of a few hundred $I\ts{sat}$, the
off-resonant atomic scattering rate can be calculated to be of the
same order of magnitude as the PA rate for a cloud at
$10^{13}$~cm$^{-3}$ density. Although the photon recoil associated
with each atomic scattering event contributes to the heating by
about twice the recoil energy (which is small compared to the
heating per molecular decay event), the scattering may leave the
atom in a different magnetic substate for which Penning ionization
is not suppressed. A subsequent Penning-ionizing collision may
cause significant heating of the cloud. On the other hand, heating
due to off-resonant scattering varies slowly with detuning
compared to heating due to PA. In a PA experiment, $I_1$ may be
chosen so as to obtain an sufficiently large rate coefficient
$K\ts{f}$ with a minimum of background signal, thus leading to
optimum S/N ratio.

Apparently, the PA signal obtained from heating or loss is related
to the strength of the PA process only in an indirect way. This
makes the calorimetric detection method less suitable for
observing FC oscillations as in Ref.~\cite{Heinzen}.

The model used here to simulate two-laser PA spectra may also be
employed for fitting experimental data. With the temperature known
from experiment, the intensity of laser 2 and the binding energy
of the $^5\Sigma_g^+,v=14$ level may be used as fit parameters.
Moreover, the He* \molgsgp\ system has good theoretical access:
radiative coupling can be computed using accurate theoretical
data, and hyperfine interaction and higher-order partial wave
contributions are absent. The two-laser PA spectroscopy proposed
in this work may thus serve as an absolute test case for existing
theoretical models.
\subsection{\label{numresOFR}Optical Feshbach resonance}
\begin{figure}[b]
\resizebox{1\columnwidth}{!}{%
  \includegraphics{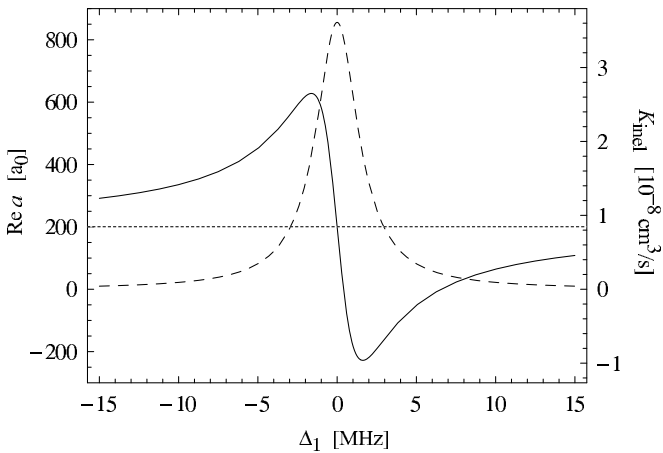}
}
\caption{\label{fig:OFR}Optical Feshbach resonance using a PA
intensity of $5\times 10^5$~W/m$^2$. Solid curve: real part of the
scattering length as a function of detuning away from a
vibrational resonance with binding energy $-250$~GHz. The natural
value of the scattering length, $a\ts{nat}$, is indicated by the
dotted horizontal line and is assumed to be $200$~a$_0$. Dashed
curve: rate coefficient for loss of atoms due to spontaneous decay
of the excited state.}
\end{figure}
An OFR requires relatively intense laser fields, which may cause
heating or loss of atoms due to off-resonant atomic scattering.
Fedichev \textit{et al.} pointed out that choosing a vibrational
level with large binding energy reduces the losses due to atomic
scattering, despite the fact that higher intensities will be
needed to achieve the same modification of
$a$~\cite{Fedichev1996a}. The purely long range states below $D_0$
discussed above have comparatively small binding energies with
respect to the $D_0$ asymptote, and the laser intensity required
for an OFR can be estimated to cause strong off-resonant atomic
scattering. The subsequent heating and losses would severely limit
the lifetime of a cold cloud of atoms, and preclude the use of the
purely long range states for an OFR.

As an alternative, the present calculation assumes the existence
of a vibrational level with a binding energy of $-250$~GHz,
located in a (\molgsgp) $^5\Sigma_u^+$ potential below $D_2$. The
$^5\Sigma_u^+$ potential couples at long-range to the Hund's case
(c) states $1_u$ and $2_u$, for which is the interaction is well
known~\cite{Venturi2003,Leonard2003b}. The short-range part of
this potential has never been considered by experiment or theory.
Nevertheless, it is to be expected that autoionization of this
state is strongly suppressed, similar to the suppression of
Penning ionization during collisions between spin-polarized He*
atoms~\cite{Leonard2004}. The long-range part of the vibrational
wavefunction is obtained by inward numerical integration of the
radial Schr\"odinger equation, assuming a binding energy of
$-250$~GHz and taking only the long-range part of the interaction
into account.
For simplicity, the long-range part is represented by a
$-C_3/R^3-C_6/R^6$ potential (with the dispersion coefficients
from Ref.~\cite{Venturi2003}), which has a negligible effect on
the wavefunction of the state bound by $-250$~GHz.
The inner part of the excited-state wavefunction thus remains
unknown. However, this part of the wavefunction oscillates rapidly
with $R$, and it can be estimated to contribute only marginally to
the FC overlap integral involved in the matrix element in
Eq.~(\ref{eq:gamma}). Now, the effect of PA light (nearly resonant
with the vibrational level) on the scattering length and the
inelastic loss rate coefficient can be studied using
Eqs.~(\ref{eq:A}), (\ref{eq:B}), and (\ref{eq:Kel}).
Fig.~\ref{fig:OFR} shows the familiar dispersive behaviour of $A$
versus detuning ($A$ may even become negative), and the resonant
character of $K\ts{inel}$. Inelastic losses stem from decay of the
excited vibrational level by spontaneous emission, which may
result either in fast and/or untrapped atoms or in autoionizing
\molgsgs\ molecules, or from autoionization of the excited state
at short internuclear range.
\begin{figure}
\resizebox{1\columnwidth}{!}{%
  \includegraphics{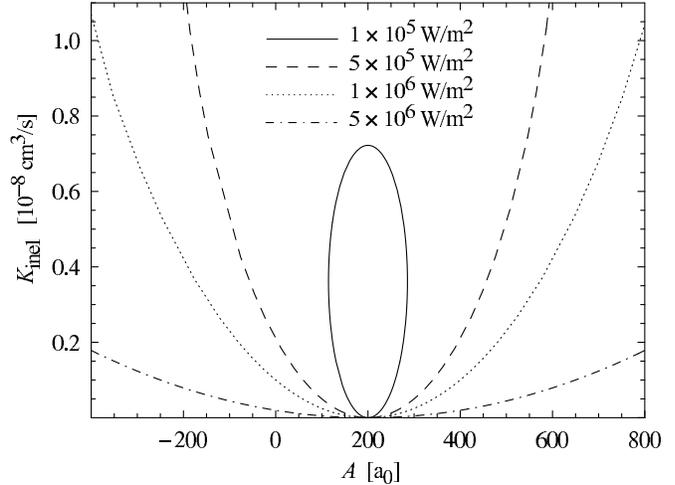}
}
\caption{\label{fig:SCVSL}Parametric plot of $K\ts{inel}$ versus
 $A$, for several PA laser
intensities and with $\Delta_1$ being varied as a parameter. Here,
$a\ts{nat}$ is assumed to be 200~a$_0$.}
\end{figure}

\begin{figure}[!]
\resizebox{1\columnwidth}{!}{%
  \includegraphics{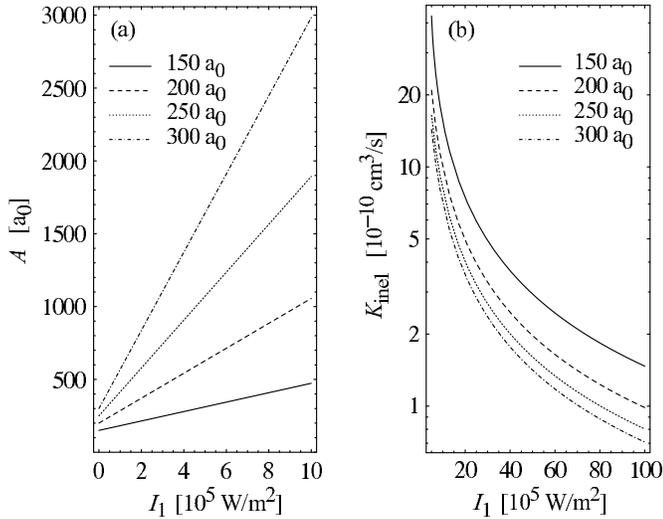}
}
\caption{\label{fig:SCVSanat}(a) Maximum achievable value for $A$
versus PA intensity $I_1$. The various curves correspond to
different values for $a\ts{nat}$. (b) Inelastic loss rate
coefficient for the case of vanishing $A$ versus PA intensity, for
several values of $a\ts{nat}$.}
\end{figure}

A more convenient graphical representation of the relation between
$A$ and $K\ts{inel}$ is shown in Fig.~\ref{fig:SCVSL}. Here, $A$
and $K\ts{inel}$ are plotted for various PA intensities, with the
detuning being varied as a parameter. It is clear that a larger
intensity allows for larger modification of the scattering length
at the same level of losses. For the particular case in which $A$
is made to vanish, the losses are seen to \emph{decrease} with
increasing PA intensity (here it should be mentioned that for
increasing PA intensity, $A$ becomes zero for increasing values of
$\Delta_1$). Another crucial role is played by the natural value
of the scattering length, $a\ts{nat}$. This is illustrated in
Fig.~\ref{fig:SCVSanat}(a), where the maximum achievable value for
$A$ is plotted as function of PA intensity for several values of
$a\ts{nat}$ lying within the uncertainty of the theoretical
prediction~\cite{dispersion}. It follows that for a larger value
of $a\ts{nat}$ the modification will be larger too. Similarly, the
parameters determining a vanishing value for $A$ depend strongly
on $a\ts{nat}$ as well. As a consequence, the inelastic loss rate
coefficient for the case $A=0$ decreases with increasing values of
$a\ts{nat}$, as can be seen in Fig.~\ref{fig:SCVSanat}(b). This
strong dependence on $a\ts{nat}$ emphasizes the importance of an
accurate (spectroscopic) determination of the scattering length.

For any of the cases discussed above, the timescale on which the
temperature and number of atoms in the cold cloud will be
seriously affected is small. For a cloud of He* atoms at a density
of 10$^{13}$~cm$^{-3}$, PA losses set a timescale of
$10^{-5}$--$10^{-3}$~s, whereas off-resonant atomic scattering
typically limits the lifetime to several milliseconds or longer.
These figures confirm the results of an earlier study of prospects
for optical Feshbach resonances in He*~\cite{Mosk2002}. In fact,
the elastic rate never exceeds the inelastic rate by far, and it
even is smaller in most of the cases discussed above. This means
that the present scheme for optical control of the scattering
length can be used only for applications in which short,
impulse-like changes of the scattering length are required. For
studies of a He* BEC in an optical lattice, which offers the
unique feature of detection of ions produced in Penning
collisions, tunability of the scattering length may be
interesting~\cite{Fedichev2004}. For instance, for the case of two
He* atoms occupying the same lattice site, the PA rate (and thus
the modification of $a$) may be enhanced~\cite{Jaksch2002}. This
would reduce the losses associated with the conditions which set
$a$ to a particular value.  Another point of further study may be
the behaviour of the Penning-collision rate as a function of $a$.
%
%

When designing an experiment to optically control the scattering
length, the binding energy of a suitable vibrational level in the
excited-state potential should be known. In the case of alkali
elements, the location of suitable bound states can often be
obtained from existing spectroscopic data on the excited-state
potential. In contrast to the case of alkali atoms, helium dimers
do not naturally exist, which explains the poor amount of
spectroscopic data on the \molgsgp\ states. Theoretical data are
available only for the (\molgsgp) $^3\Sigma_g^+$ state; data of
singlet and quintet molecular potential energy curves have
hitherto not been published. To avoid an extensive spectroscopic
search for lines near the desired binding energy, the
accumulated-phase method might be employed to assist in predicting
their locations~\cite{Moerdijk1994,Leonard2004}. For an accurate
prediction of vibrational level positions, however, the
availability of \emph{ab initio} data is probably crucial.

When deciding on a suitable vibrational level to use for an
optical Feshbach resonance, a point of concern is whether nearby
molecular vibrational resonances in singlet or triplet potentials
can be excited by the PA light. Such resonances, which will be
broadened to almost the vibrational level spacing because of the
large (near unity) probability of autoionization at short
range~\cite{Mueller1991}, introduce additional losses. However,
for the Condon radius corresponding to excitation at $-250$~GHz,
all nearby molecular potentials are nearly pure Hund's case (a)
states, as the relativistic atomic interactions that mix them at
long range are small compared to the resonant dipole-dipole
interaction. Starting from a sample of spin-polarized He* atoms,
this means that only quintet \textit{ungerade} states can be
accessed, and no PA to molecular states coupling to strongly
autoionizing singlet and triplet configurations is possible.
%
%
\section{\label{conclusion}Conclusion}
Two-laser PA spectroscopy of ultracold spin-polarized He* atoms,
aiming at the detection of the least-bound vibrational level in
the $^5\Sigma_g^+$ scattering potential, has been investigated
numerically. The binding energy of this state may be used to infer
an accurate value for the He* $s$-wave scattering length. The
scheme, which assumes the purely long range $v=0$ vibrational
level in the $0_u^+$ potential well below $D_0$ as the
intermediate level, may result in a detectable two-photon
stimulated Raman transition for feasible experimental parameters.
Different approaches to observe the two-laser transition are
discussed within the context of a calorimetric detection
method~\cite{Leonard2003a,Leonard2003b}.

In addition, optical control of the He* scattering length using
optical Feshbach resonances has been explored numerically.
Modification of the scattering length to very large
($>10^3$~a$_0$) values or to vanishing or even negative values is
anticipated for experimentally realizable conditions, although the
magnitude of the optical Feshbach resonance depends strongly on
the natural value of the scattering length. Also, this method to
control the scattering length may be applied only for a very short
($10^{-5}$--$10^{-3}$~s) duration, to avoid substantial loss
and/or heating of trapped atoms. This will undoubtedly complicate
a possible experimental implementation of an OFR. Notwithstanding
all the experimental disadvantages as compared to magnetic
Feshbach resonances, optical Feshbach resonances are presumably
the only handle one has to modify the value of the He* scattering
length. Finally, several experimental issues concerning the search
for and choice of a suitable excited-state vibrational level have
been discussed.
\section{Acknowledgments}
Allard Mosk is gratefully acknowledged for useful comments on the
manuscript and fruitful discussions. The authors further wish to
thank the members of the ultracold metastable helium team at
\'Ecole Normale Sup\'erieure (ENS), J\'er\'emie L\'eonard, Olivier
Dulieu, and Servaas Kokkelmans for stimulating conversations. The
stay of J. K. at ENS was financially supported by means of a Marie
Curie European fellowship.
\section{\label{appendix}Appendix: matrix elements of the radiative coupling}
The matrix element in Eq.~(\ref{eq:gamma}) involves the
energy-norma\-lized state $\ket{f_E}$ (the label $l$ is dropped
since only the $l=0$ contribution is taken into account), which
can be written as the product of a radial wavefunction, a
rotationless angular part, and the rotational part of the
molecular state:
\begin{equation}
\ket{f_E}=\frac{u_{f,E}(R)}{R}\ket{^5\Sigma_g^+}\ket{J\Omega M}.
\end{equation}
Here, $\Omega$ is the projection of $J$ on the internuclear axis,
whereas $M$ is the projection of $J$ on the lab-fixed $z$-axis,
defined by the field of the magnetic trap. For spin-polarized He*
atoms, $M=2$; hence $J=2$ (for $l=0$). $\Omega$ depends on the
alignment of the internuclear axis with respect to the $z$-axis.
The energy-normalized radial wavefunction, $u_{f,E}(R)$, is
obtained by solving the Schr\"odinger equation for two free atoms
in the scattering potential with an asymptotic energy $E$, with
the long-range boundary condition~\cite{Bohn1999}
\begin{equation}\label{eq:energynorm}
u_{f,E}(R)\sim\sqrt{\frac{2\mu}{\pi\hbar^2
k}}\sin[kR+\eta(k)],\qquad R\rightarrow\infty.
\end{equation}
Here $k$ is the initial wave number, and $\eta(k)$ denotes an
energy-dependent phase shift. The left-hand side of
Eq.~(\ref{eq:energynorm}) corresponds to an energy-normalized,
free-particle state. The excited bound state can be written as
\begin{eqnarray}\label{eq:expansion}
\ket{e} &=& \frac{u_{v=0}(R)}{R}\ket{0_u^+}\ket{J'\Omega'
M'}\nonumber \\
&=& \frac{u_{v=0}(R)}{R}\{ c_1(R)\ket{^5\Sigma_u^+}
+c_2(R)\ket{^5\Pi_u}
\nonumber\\
&&+\ket{\rm{n.c.}}\}\ket{J'\Omega'M'},
\end{eqnarray}
where $c_i(R)$ are the expansion coefficients of $\ket{0_u^+}$
expressed in \emph{ungerade} Hund's case (a) states.
$\ket{\rm{n.c.}}$ is the sum of all singlet and triplet states in
the expansion $\ket{0_u^+}$, which are not coupled to the
$^5\Sigma_g^+$ lower state by a dipole transition. Therefore,
these states may be omitted in the evaluation of the matrix
element of $V\ts{rad}$. Bose statistics furthermore require that
$J'=1$. Consequently, because of dipole selection rules, only the
$M'=1$ state can be excited starting from the $M=2$ lower state,
which implies $\sigma^-$ transitions induced by laser 1.

The matrix element involves a spatial integration, which can be
split up in an angular part and a radial part (the so-called
\emph{Franck-Condon integral}). It should be noted that the
Franck-Condon \emph{principle} (i.e. approximating the integral by
replacing it with the integrand evaluated at the Condon radius)
does not apply to the transition to the $v=0$ excited state in
$0_u^+$. This is due to the fact that the $v=0$ radial
wavefunction, unlike more highly excited vibrational
wavefunctions, has a broad amplitude maximum which spans the
entire classically allowed region. Also, a common approximation in
which only the radial wavefunctions $u(R)$ in the FC integral are
included is inadequate, since the angular part of the matrix
element also depends on $R$ via the expansion coefficients
$c_i(R)$ in Eq.~(\ref{eq:expansion}). The FC integral is therefore
evaluated over the full internuclear range of overlap, including
the radial wavefunctions and expansion coefficients $c_i(R)$, with
each coefficient weighted by the value of the corresponding
angular part of the matrix element. The latter is evaluated using
the direction cosine matrix~\cite{Hougen1973,LefebvreField}. It
should also be noted that the atomic dipole moment in
Eq.~(\ref{eq:Vrad}) is modified so as to take the presence of the
magnetic field (which lifts the degeneracy of the lower state)
into account.

For the matrix element describing the radiative coupling in the
OFR, the radial integration is limited to the internuclear range
for which the excited-state wavefunction is known
(Section~\ref{numresOFR}), thereby neglecting the short-range
contribution. Since in this case the excited state is virtually a
pure Hund's case (a) state, the radial integral only contains the
radial wavefunctions of the lower and excited states.

The bound-bound matrix element in Eq.~(\ref{eq:Rabi}) is obtained
along the same lines as given above, but now involving an
$M'=1\rightarrow M=2$ transition. It is furthermore assumed that
the magnetic moment of the $v=14$ bound state in the \qsgp\
potential is identical to that of the initial free-atom pair, such
that all Zeeman shifts due to the field of the magnetic trap
cancel out.
%
%
%

\begin{thebibliography}{}
\bibitem{dispersion}In the calculation, $a\ts{nat}$ is varied by slight modification
of the dispersive long-range part of the $^5\Sigma_g^+$ scattering
potential.
\end{thebibliography}
%

\end{document}